\documentclass[manuscript]{aastex}
\begin{document}

\title {Emission Patterns and Light Curves of Gamma-Rays in the Pulsar Magnetosphere with a Current-Induced Magnetic Field}
\author{X. Li  \& L. Zhang}
\affil{Department of Physics, Yunnan University, Kunming, China}
\email{lizhang@ynu.edu.cn}

\begin{abstract}
We study the emission patterns and light curves of gamma-rays in
the pulsar magnetosphere with a current-induced magnetic field
perturbation. Based on the solution of a static dipole with the
magnetic field induced by some currents (perturbation field), we
derive the solutions of a static as well as a retarded dipole with
the perturbation field in the Cartesian coordinates. The static
(retarded) magnetic field can be expressed as the sum of pure
static (retarded) dipolar magnetic field and the static (retarded)
perturbation field. We use the solution of the retarded magnetic
field to investigate the influence of the perturbation field on
the emission patterns and light curves, and we apply the perturbed
solutions to calculate the gamma-ray light curves for the case of
the Vela pulsar. We find out that the perturbation field induced
by the currents will change the emission patterns and then light
curves of gamma-rays, especially for a larger perturbation field.
Our results indicate that the perturbation field created by the
outward-flowing (inward-flowing) electrons (positrons) can
decrease the rotation effect on the magnetosphere and makes
emission pattern appear to be more smooth relative to that of the
pure retarded dipole, but the perturbation field created by the
outward-flowing (inward-flowing) positrons (electrons) can make
the emission pattern less smooth.
\end{abstract}


\section{Introduction}

Since pulsed gamma-ray emission from a large number of pulsars by
the Large Area Telescope (LAT) on the Fermi Gamma-Ray Space
Telescope have been discovered \citep{Abdo10}, much information
are available for constraining pulsar magnetosphere geometry.
Rotation-powered pulsars are widely believed to have their
magnetospheres in which charged particles are accelerated to
relativistic energy and generate multiband spectra of pulsed
emission. In order to simulate pulsar magnetosphere, various
approximations have been proposed. Vacuum magnetosphere model is
analytic model for the electromagnetic field of a rotating
magnetic dipole \citep{Deutsch55}, but it is not an appropriate
physical model of an active pulsar magnetosphere filled with
charges and currents. Force-free magnetosphere is probably a
closer approximation to a real pulsar than the vacuum solution,
however it is not a truly self-consistent model since the
production of pair plasma requires particle acceleration and thus
a break down of force-free conditions in some regions of the
magnetosphere. At present, there are two kinds of three
dimensional (3D) emission models: one is based on the geometrical
consideration such as two pole  caustic (TPC) model in the frame
of a retarded dipole \citep[e.g.,][]{DR03,DHR04,FZ10} and annular
gap (AG) model in the frame of free-force approximation
\citep{BS10a}; another is based on both physical and geometrical
consideration such as slot gap (SG) \citep{HSDF08} and outer gap
models
\citep[e.g.,][]{RY95,CRZ00,ZC00,ZC01,ZC02,Tet08,ZL09,W09,LZ10,WTC11}.

How we can describe a pulsar magnetosphere filled with charges and
currents? Since the currents will induce magnetic field
perturbation, resulting to a distorted magnetic field relative to
dipole field, \citet{MH09} described an approximate perturbation
field for a pair-starved current flow in the open zone in the
frame of static magnetic dipole. In their treatment, the magnetic
field is described as $\textbf{B}=\textbf{B}^{\rm (d)} + \epsilon
\textbf{B}^{(1)}$, where $\textbf{B}^{\rm (d)}$ is a pure dipole
magnetic field anchored into the neutron star (NS), $\epsilon$ is
the perturbation amplitude which determines the strength of the
$\textbf{B}^{(1)}$ relative to the dipole one , and
$\textbf{B}^{(1)}$ is the magnetic field generated by the
self-consistent electric currents in the domain of the
magnetosphere with open field lines, where a singularity exists on
the symmetry axis (details see in \citet{MH09}). However, although
basic features of the pulsar magnetosphere in a static dipolar
magnetic field approximation can be understood, it is more
realistic that the magnetosphere be approximated as a rotating
inclined magnetic dipole. The magnetic field lines of a rotating
inclined dipole are especially different from those of a static
dipole for large inclination angles. \citet{RW10} investigated the
magnetic field structure with a current-induced field in the frame
of the retarded magnetic dipole. To apply the solution of
\citet{MH09} to the retarded dipole, \citet{RW10} made following
treatments: (1) shifted the magnetic axis ($r$, $\theta_{\rm
B}=0$, $\phi_{\rm B}$) to match that of the swept back dipole
(i.e. mapped the magnetic axis line onto the swept back curve) and
(2) exponentially tapered the singularity appeared on the symmetry
axis so that the field lines can be integrated to determine the
last closed field line surface. Obviously, such a treatment method
is highly simplified.

In this paper, we study the emission patterns and light curves in
the pulsar magnetosphere with a current-induced magnetic field
perturbation. We derive the analytic solution of the retarded
magnetic field with a current-induced magnetic field{\bf ;} this
solution is that of \citet{MH09} when rotating effects are
ignored. We find that our calculated results are different from
those given by \citet{RW10}, for example open zone boundary (polar
cap) foot-points. Using such a solution, we calculate emission
patterns and light curves for the Vela pulsar with different
inclination angels in the frame of two pole outer gap model, where
a self-consistent treatment is that both simulation of the
magnetosphere and relativistic effects are performed in the
inertial observer's frame (IOF) \citep[e.g.,][]{TCC07,BS10b}. The
paper is organized as follows. In \S 2, the solution of the
magnetic field structure with current-induced magnetic field is
given. Various emission patterns and light curves are calculated
in \S 3. Finally, a brief discussion and conclusions are given in
\S 4.

\section{Magnetic Field Structure}
\label{sec:model}

In a pulsar magnetosphere, radiating charges accelerated in gaps
will create some currents in the open zone. If gap-closing pair
front produce densities comparable to the co-rotation value for at
least some of the open field lines, this current will induce a
magnetic field \citep[e.g.,][]{MH09,RW10}. For a static dipole, the
local magnetic field can be described as \citep{MH09}
\begin{equation}
\textbf{B}=\textbf{B}^{\rm (d)} + \epsilon \textbf{B}^{(1)}\;,
\label{Bfield}
\end{equation}
where $\textbf{B}^{\rm (d)}$ is the pure dipole magnetic field and
is given by
\begin{equation}
\textbf{B}^{\rm (d)}= \frac{1}{r^3}\left[3({\bf m}\cdot{\bf \hat
r}){\bf \hat r}-{\bf m} \right]\;, \label{BSDfield}
\end{equation}
where $\bf m$ is the magnetic moment and its three components in
Cartesian coordinates can be expressed as (the derivation is shown
in Appendix A)
\begin{eqnarray}\label{ms}
m_x&=&\frac{r}{2}\left[(3x^2-2r^2)B^{\rm (d)}_x+3xyB^{\rm (d)}_y+3xzB^{\rm (d)}_z\right]\nonumber\\
m_y&=&\frac{r}{2}\left[3yxB^{\rm (d)}_x+(3y^2-2r^2)B^{\rm (d)}_y+yzB^{\rm (d)}_z\right]\\
m_z&=&\frac{r}{2}\left[3xzB^{\rm (d)}_x+3yzB^{\rm
(d)}_y+(3z^2-2r^2)B^{\rm (d)}_z\right]\nonumber \;, \label{Mxyz}
\end{eqnarray}
where $r=\sqrt{x^2+y^2+z^2}$, and $B^{\rm (d)}_x$, $B^{\rm (d)}_y$,
and $B^{\rm (d)}_z$ are three
components of the magnetic field given by Eq. (\ref{BSDfield}). In
Eq. (\ref{Bfield}), $\epsilon$ is the perturbation amplitude, and
$\textbf{B}^{(1)}$ is the magnetic field generated by the
self-consistent electric currents in the domain of the magnetosphere
with open field lines. After introducing cosine of the angle between
the neutron star(NS) rotation axis and radius-vector of a given
point $s=\cos\alpha\cos\theta+\sin\alpha\sin\theta\cos\phi$ and its
derivatives over $\theta$ and $\phi$, \citet{MH09} derived the
analytic expressions of Eq. (\ref{Bfield}) in magnetic polar
coordinates for the static dipole, where $s = 1$ determines the
symmetry axis in magnetic coordinates which is the NS rotation axis.
In Appendix A, we change expressions ($B_r$, $B_{\theta}$,
$B_{\phi}$) of Eq. (\ref{Bfield}) in magnetic polar coordinates to
those ($B_x$, $B_y$, $B_z$) in Cartesian coordinates, and then give
three components of the magnetic moment $\bf{m'}$ in which the
magnetic field is given by Eq. (\ref{Bfield}) (i.e. the perturbation
field is included) by changing ($B^{\rm (d)}_x$, $B^{\rm (d)}_y$,
$B^{\rm (d)}_z$) to ($B_x$, $B_y$, $B_z$) in Eq. (3). In such a
treatment, the magnetic moment $\bf{m}$ in the pure dipole magnetic
field changes to $\bf{m'}$ in the magnetic field given by Eq.
(\ref{Bfield}). We call $\bf{m'}$ as an effective magnetic moment.
 The difference between $\bf{m}$ and $\bf{m'}$ depends on the
perturbation amplitude $\epsilon$, \emph{$\delta \bf{m}=|\bf{m} -
\bf{m'}|=0$} when $\epsilon=0$.

Since the perturbation field becomes singular along the rotation
axis where $s\rightarrow \pm1$, \citet{RW10} exponentially tapered
this singularity ($\propto 1-\exp[(|s|-1)/\sigma_s]$, with $\sigma_s
= 0.25$) so that the field lines can be integrated to determine the
last closed field line surface. Here we will use this method.

\begin{figure}
\epsscale{1} \plotone{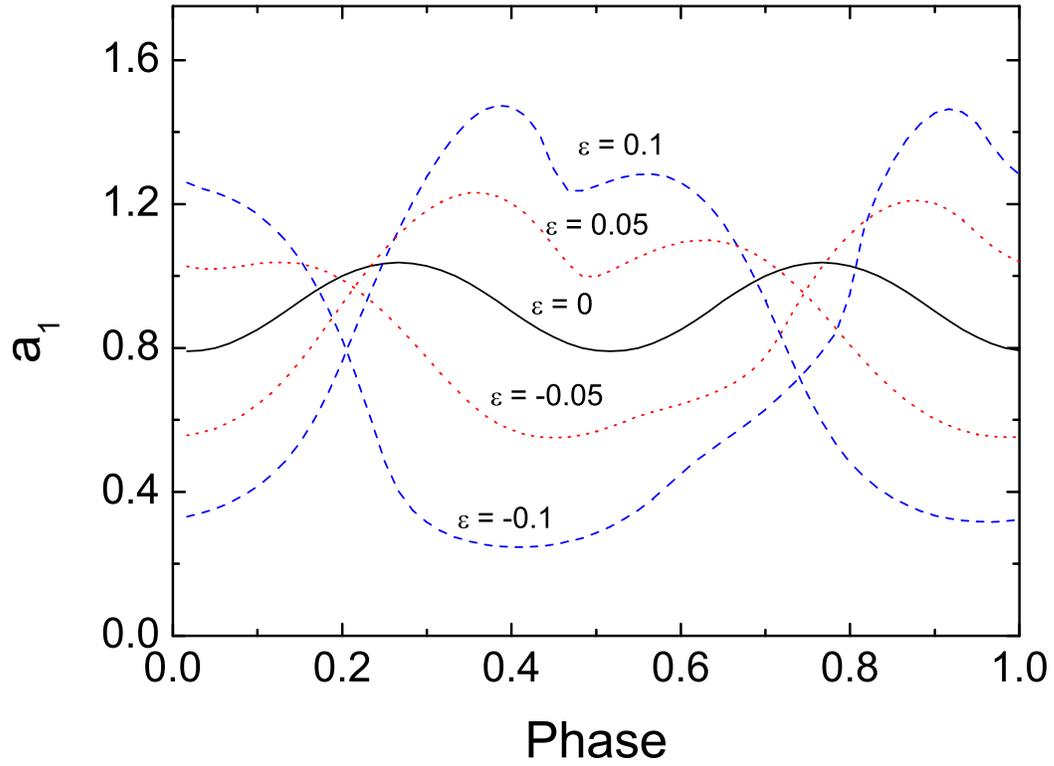} \caption{\label{Fig1}Changes of
open zone boundary (polar cap) foot-points with phase for the
$\alpha = 70^{\circ}$ static dipole with current-induced
perturbations $\epsilon=0, \pm0.05$, and $\pm 0.1$. The Vela
pulsar's parameters are used. }
%\end{center}
\end{figure}

\begin{figure}
\epsscale{1.0} \plotone{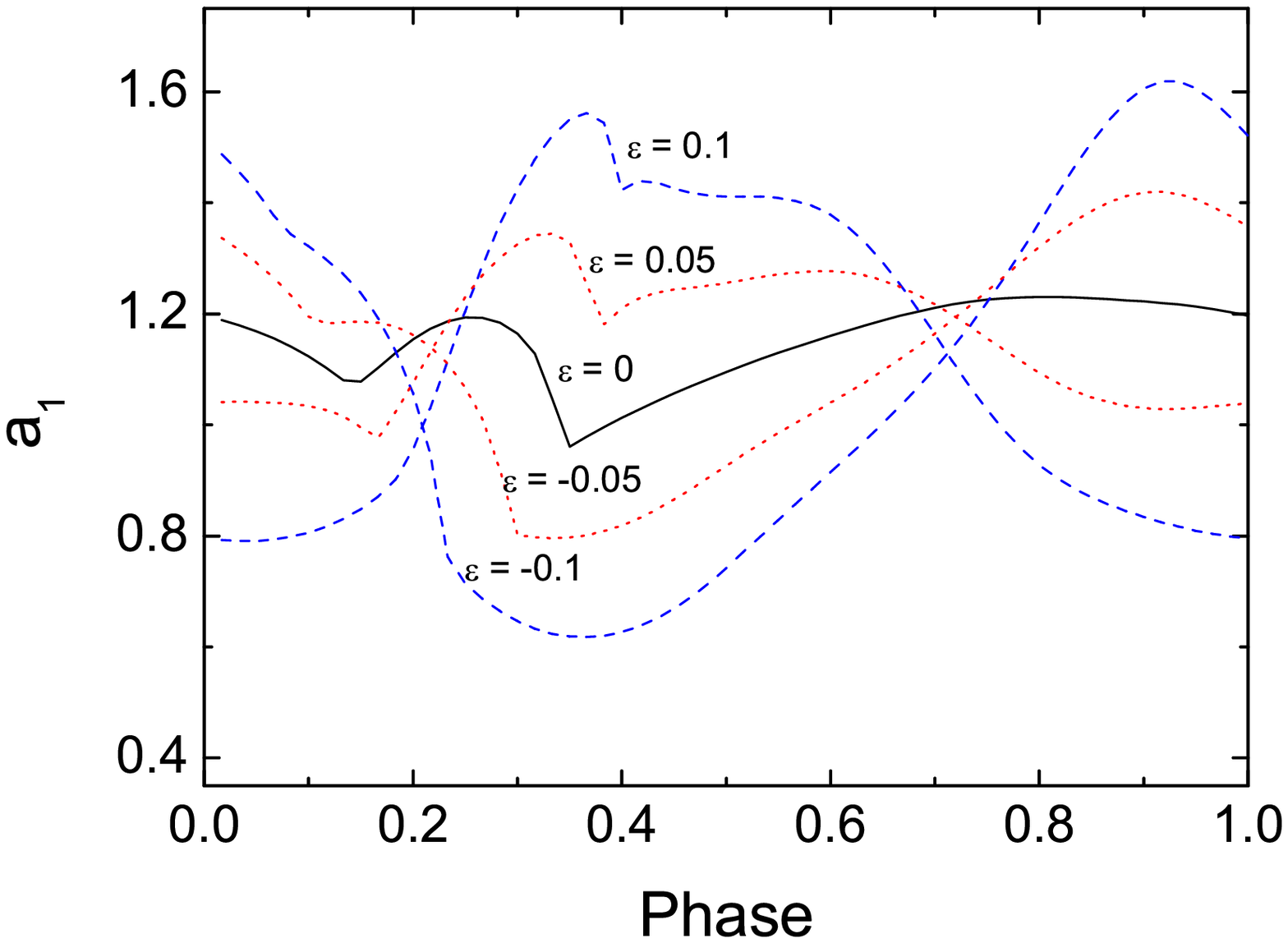} \caption{\label{Fig2}Changes of
open zone boundary (polar cap) foot-points with phase for the
$\alpha = 70^{\circ}$ retarded dipole with current-induced
perturbations $\epsilon=0, \pm0.05$, and $\pm 0.1$. The Vela
pulsar's parameters are used. }
%\end{center}
\end{figure}

\begin{figure}
\epsscale{1.0} \plotone{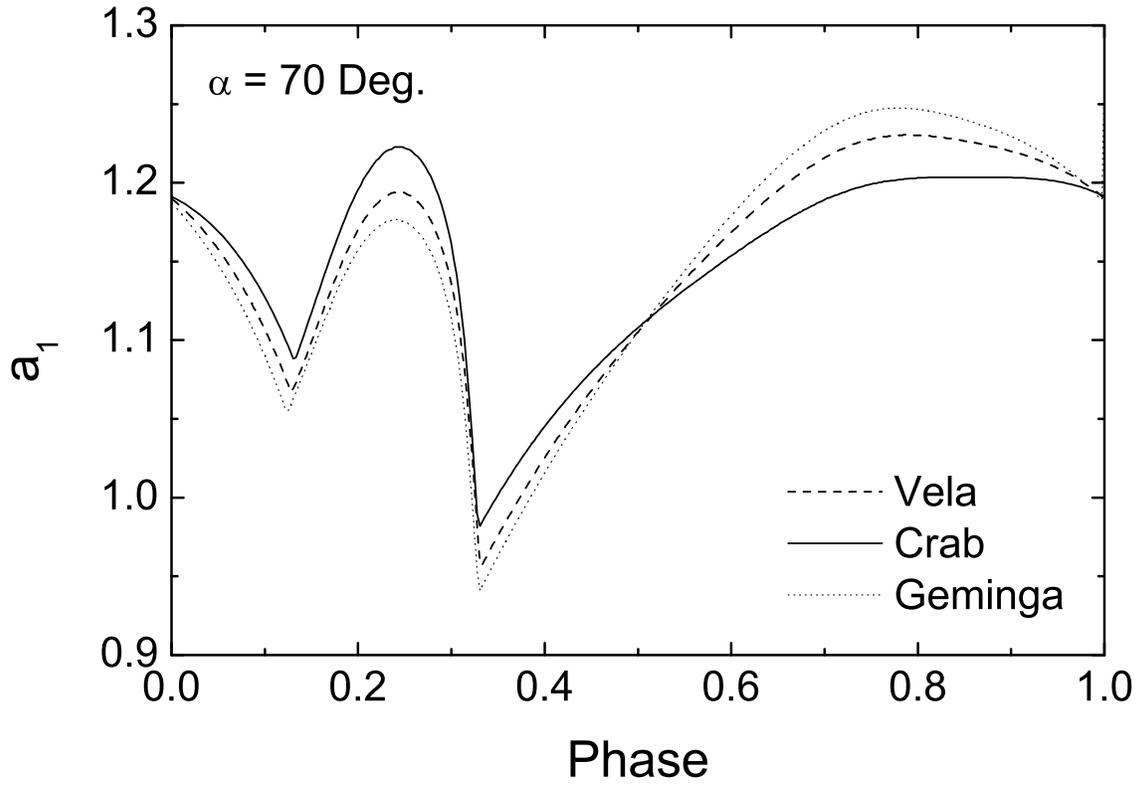} \caption{\label{Fig3}Changes of
open zone boundary (polar cap) foot-points with phase for the
$\alpha = 70^{\circ}$ retarded dipole with $\epsilon=0$ for the
periods of Crab, Vela, and Geminga pulsars.  }
%\end{center}
\end{figure}

\begin{figure}
\epsscale{0.7} \plotone{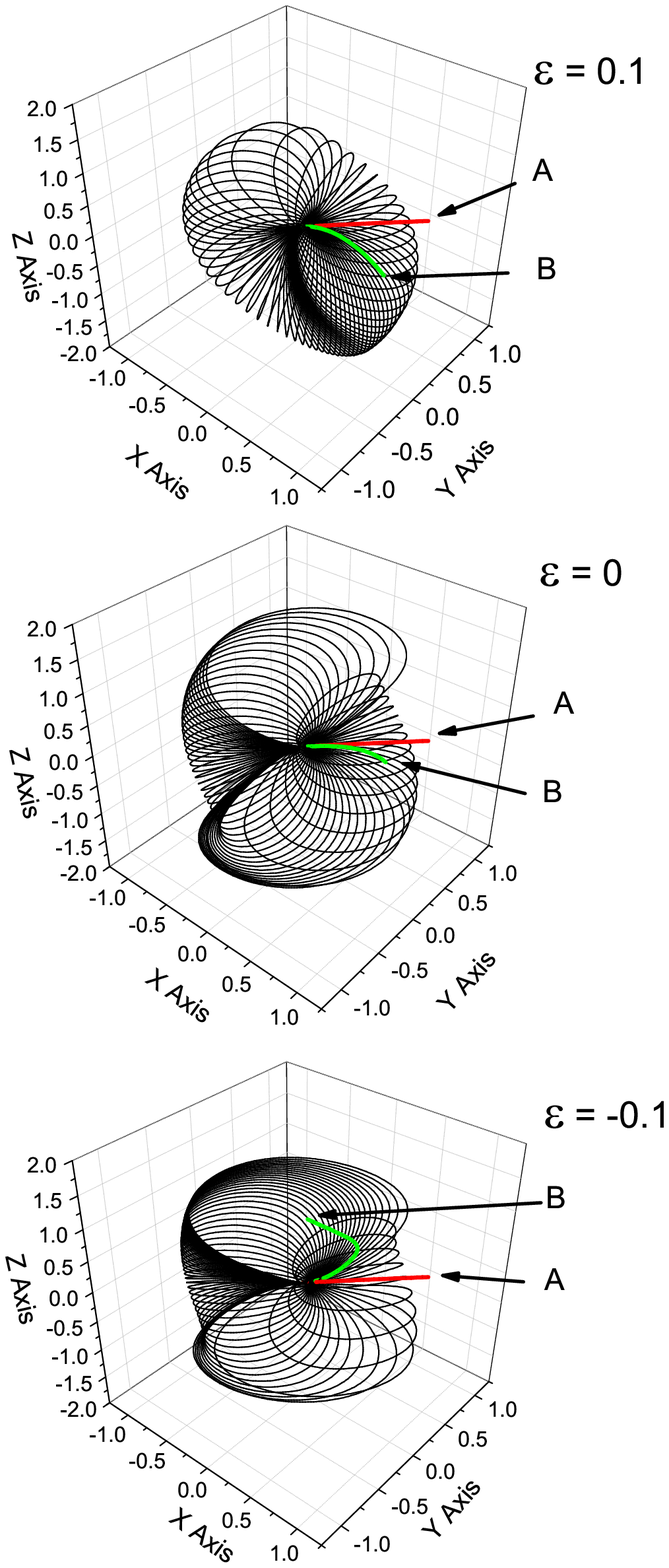} \caption{\label{Fig4}A
three-dimensional view of last closed field lines for a retarded
dipole with or without the perturbation field inclined to the
rotation axis with angle $\alpha=50^{\circ}$ for Vela pulsar. The
magnetic field is given by Eq. (\ref{Br}). The panels from top to
bottom represent that the cases of $\epsilon=0.1$, $\epsilon=0.0$,
and $\epsilon=-0.1$. For each panel, the thick red line marked by
A represents the magnetic axis of pure static dipole, and the
thick green curved line marked by B represents the magnetic axis
of the retarded magnetic field given by Eq.(\ref{Br}). }
%\end{center}
\end{figure}

The structure of the magnetosphere can still only be described by
a specific features such as the last closed field lines. The
footprints of the last closed field lines on the stellar surface
define the polar cap shape. In the static approximation, the polar
cap shape is a circle with radius of  $ R_{\rm pc}= R_0(
R_0/R_{\rm L})^{1/2}$ for an aligned dipole, where $R_{\rm
L}\equiv c\Omega^{-1}$ is the light cylinder radius, and $R_0$ is
radius of the neutron star. we define the initial polar cap's edge
using $(x_0,y_0,z_0)=[R_{\rm pc}{\rm cos}(\phi_{\rm p}),R_{\rm
pc}{\rm sin}(\phi_{\rm p}),(R_0^2-R_{\rm pc}^2)\ ],$ where
$\phi_{\rm p}$ is the azimuthal angle about the magnetic axis.
Then we try to find the scaling factors $a_0$ such that
$(x_0',y_0',z_0')=[a_0x_0,a_0y_0,(R_0^2-a_0^2R_{\rm p}^2)^{1/2}\
]$ by using the Runge-Kutta method \citep{ZC01}. Next we define
another scaling factor $a_1$ to denote the footprints of the open
field lines as
$(x,y,z)=(a_1x_0',a_1y_0',[R_0^2-(x^2+y^2)]^{1/2})$. As an
example, we show the variation of open zone boundary (polar cap)
foot-points with phase for the $\alpha = 70^{\circ}$ static dipole
with current-induced perturbations $\epsilon=0, \pm0.05$, and $\pm
0.1$ for Vela pulsar parameters (i.e. its period is $P=0.0893$ s
and surface magnetic field is  $B=3.3\times 10^{12}$ G) in Fig.
\ref{Fig1}. It can be seen that the open zone boundary appears two
symmetric peaks if the current-induced B perturbation is not
included (i.e. $\epsilon=0$). However, such a symmetry breaks when
the current-induced B perturbations is included (i.e.
$\epsilon\neq0$).

For an inclined dipole of constant magnitude, rotating with
angular frequency $\Omega$ about the $\hat{z}$-axis in which the
magnetic field consists of pure retarded dipolar magnetic field
and perturbation field, the magnetic moment is
\begin{equation}
\bf{m}^{'r}=\bf{m}^{r} + \epsilon \bf{m}^{r}_{p}\;,
\end{equation}
where $\bf{m}^{r}$ and $\bf{m}^{r}_{p}$ are the magnetic moments
of pure retarded dipolar magnetic field $\bf{B}^{r}$  and
perturbation magnetic field $\bf{B}^{r}_{p}$, respectively and the
expression are given in Appendix B. In such a case, the magnetic
field is
\begin{equation}
\bf{B}^{'r}=\bf{B}^{r} + \epsilon \bf{B}^{r}_{p}\;, \label{Br}
\end{equation}
where the expressions of  $\bf{B}^{r}$ and $\bf{B}^{r}_{p}$ are
given in Appendix B.

In Fig. \ref{Fig2}, we show the variation of open zone boundary
foot-points with phase for the $\alpha = 70^{\circ}$ retarded
dipole with current-induced perturbations $\epsilon=0, \pm0.05$,
and $\pm 0.1$ for comparison with the results of \citet{RW10}.
Since the rotating effect in the retarded dipole is included, two
symmetric peaks disappear even the current-induced B perturbation
is not included (i.e. $\epsilon=0$). This result with $\epsilon=0$
is consistent with that of  \citet{RW10} (see their top panel of
Fig. 2), but there is a small difference between our result and
the result of \citet{RW10}. In fact, the shape of the open zone
boundary foot-points depends on not only the magnetic inclination
angle but also the pulsar's period, we show the shapes for the
periods of Crab, Vela, and Geminga with $\alpha=70^{\circ}$ in
Fig. \ref{Fig3} to account for this fact. Moreover, the
cylindrical radius $r_\perp=1.2R_{\rm L}=1.2c/\Omega$ was used in
\citet{RW10}, but $r_\perp=R_{\rm L}$ is used in our treatment,
leading to the difference. With increasing B perturbation (i.e.
increasing $\epsilon$), open zone boundary is significantly
different from that with $\epsilon=0$, as shown in Fig \ref
{Fig2}. More importantly, for the singularity in the pair-starved
field of \citet{MH09}, we use the smooth method of \citet{RW10} to
deal with it in static dipole field (see Fig. \ref{Fig1}), so this
singularity  does not exist in retarded dipole field. However,
\citet{RW10} took the solution of \citet{MH09}, and shifted the
magnetic axis to match the retarded dipole by simply mapped the
magnetic axis line onto the swept back curve. Therefore, our
results with $\epsilon\neq0$ are very different from those given
by \citet{RW10}. Physically, the perturbation fields with
$\epsilon<0$ and $\epsilon >0$ should have opposite roles on pure
retarded dipole field, our results show this property, so we
believe that our results are reasonable.

In the retarded dipole, the magnetic axis is curved relative to
that of pure static dipole. In Fig. \ref{Fig4}, we show a
three-dimensional view of last closed field lines for a retarded
dipole with or without the perturbation field inclined to the
rotation axis with angle $\alpha=50^{\circ}$, where the thick red
line marked by A represents the magnetic axis of pure static
dipole, and the thick green curved line marked by B represents the
magnetic axis of the retarded dipole with the perturbation field
($\epsilon\neq 0$). It can be seen that the perturbation field has
an important role in both the shape of the last closed field lines
and the magnetic axis. For example, when $\epsilon>0$, the shape
of the last closed field lines becomes more smooth and the
magnetic axis is more curved on the side of the magnetic axis of
pure retarded dipole (see top panel of Fig. \ref{Fig4}).

\section{Emission Patterns and Light Curves}

\begin{figure*}
\epsscale{1.0} \plotone{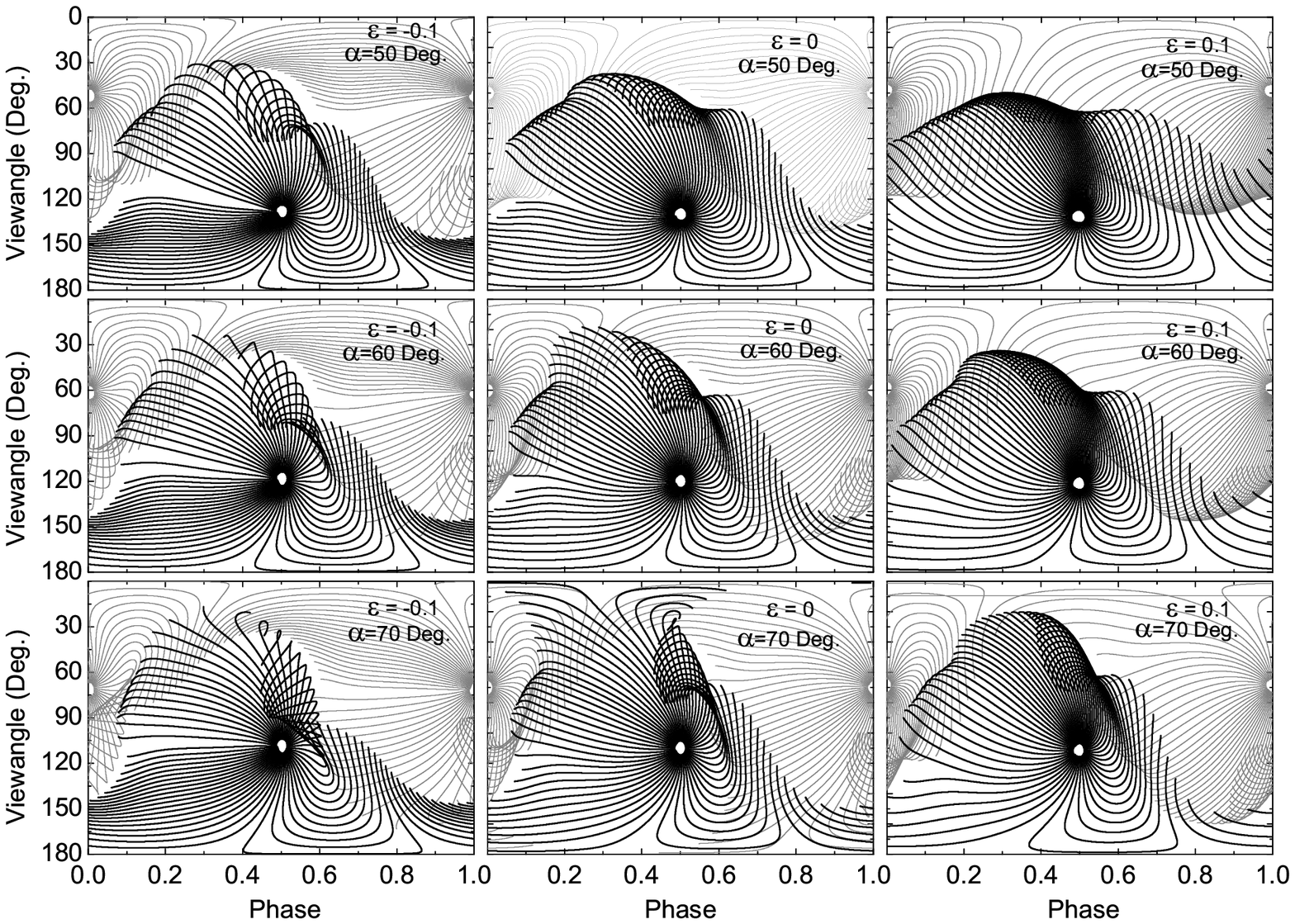} \caption{\label{Fig5} Emission
projections onto the $(\zeta , \Phi)$-plane for various magnetic
inclination angles and for $\epsilon=0.1$, 0, and -0.1. The black
lines represent the emission from one pole and the grey lines
represent the emission from another pole. Vela pulsar parameters
are used.}
%\end{center}
\end{figure*}

\begin{figure}
\epsscale{1.0} \plotone{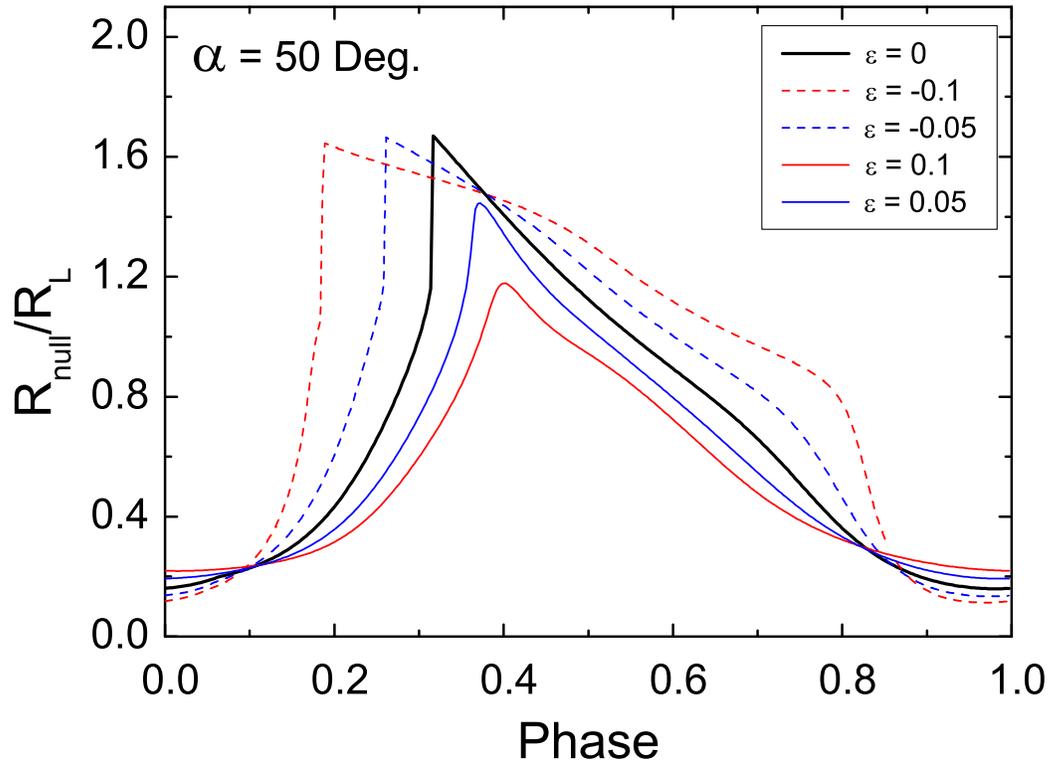} \caption{\label{Fig6} Radial
distances to the null charge surface for the last open field lines
for $\alpha=50^{\circ}$ and different values of $\epsilon$. The Vela
pulsar's parameters are used. }
%\end{center}
\end{figure}

\begin{figure}
\epsscale{0.5} \plotone{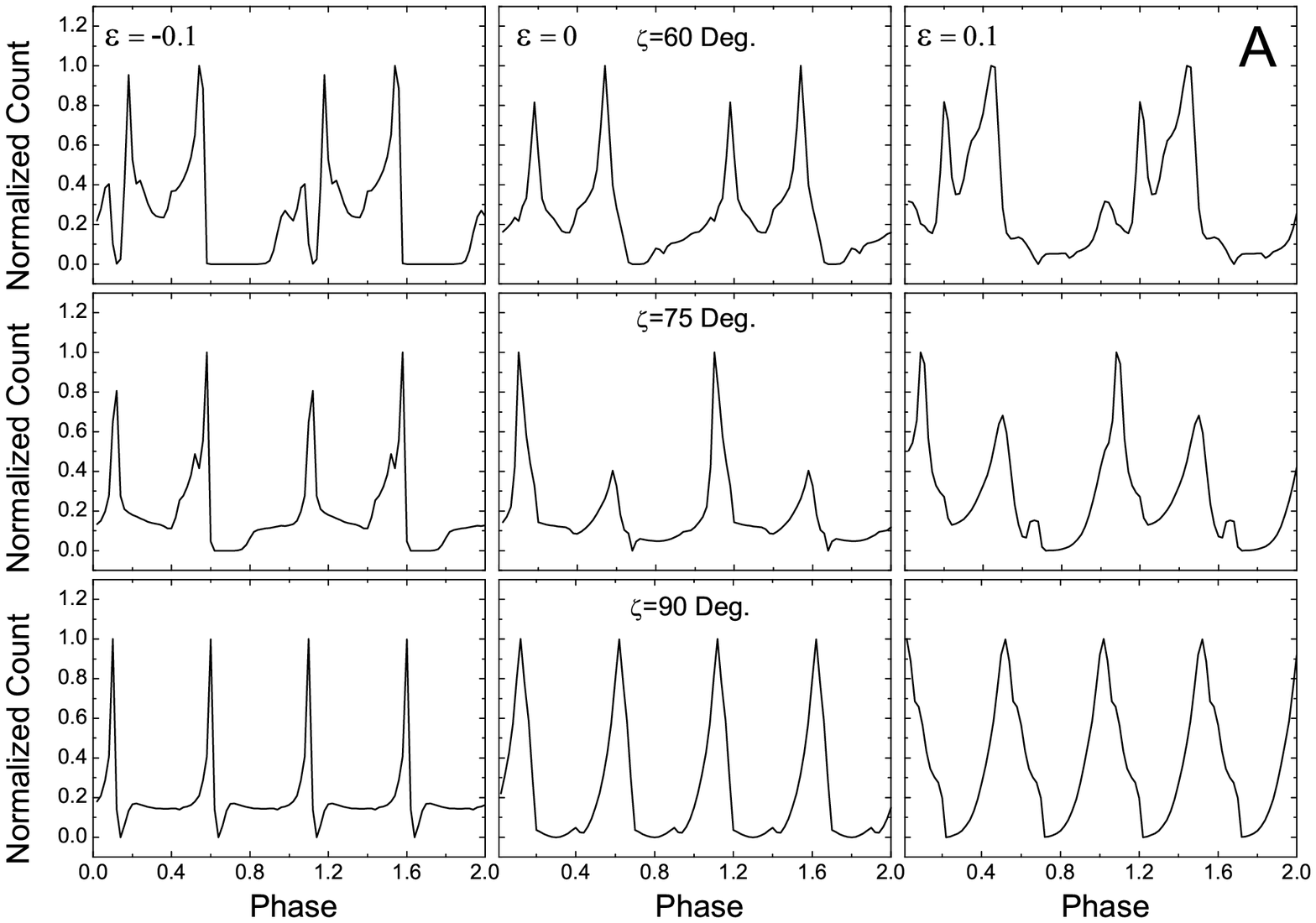} \plotone{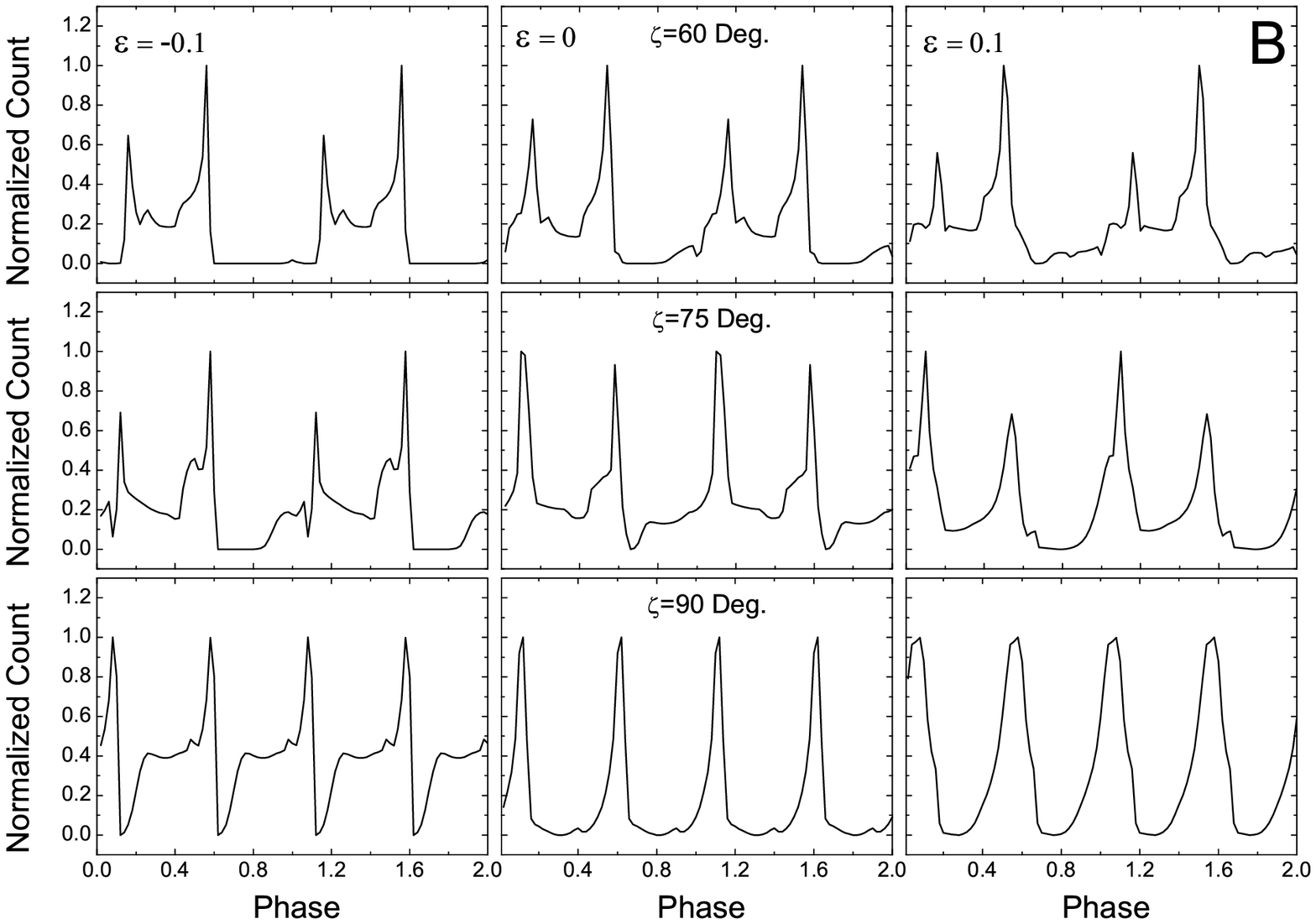}
\plotone{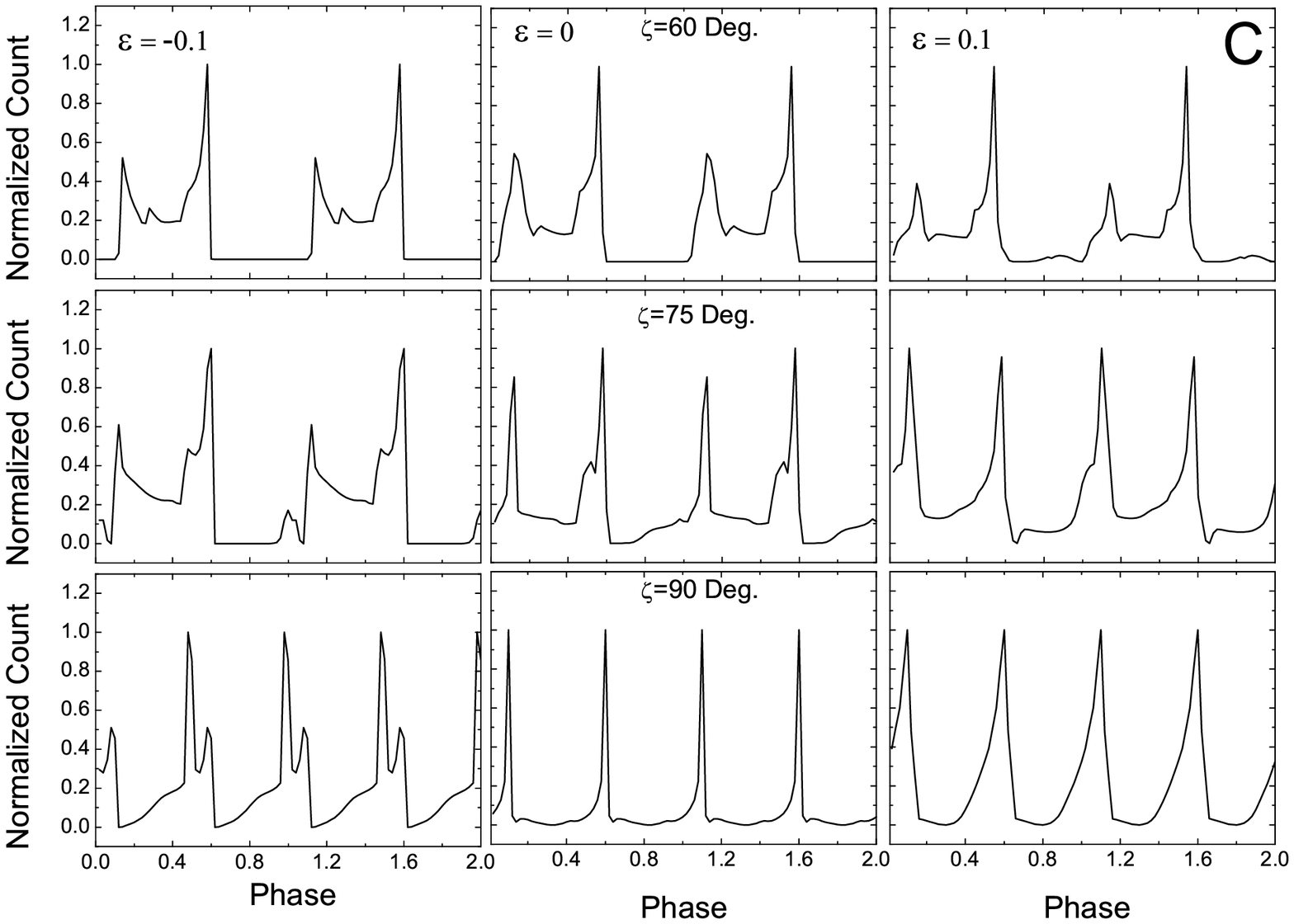} \caption{\label{Fig7} Light Curves for
different magnetic inclinations, view angles, and $\epsilon$.
Panel A: $\alpha=50^{\circ}$, $\zeta=60^{\circ}$, $75^{\circ}$,
$90^{\circ}$, and $\epsilon=-0.1$, 0, 0.1. Panel B:
$\alpha=60^{\circ}$, $\zeta=60^{\circ}$, $75^{\circ}$,
$90^{\circ}$, and $\epsilon=-0.1$, 0, 0.1. Panel C:
$\alpha=70^{\circ}$, $\zeta=60^{\circ}$, $75^{\circ}$,
$90^{\circ}$, and $\epsilon=-0.1$, 0, 0.1. Both outward emission
from one pole and inward emission from another pole are included
and the azimuthal range of outer gaps is limited by using the
method of Fang \& Zhang (2010). The Vela
pulsar's parameters are used. }
%\end{center}
\end{figure}

In order to determine photon emission from the outer gaps, 3D
pulsar's magnetosphere should be simulated and relativistic
effects including photon abberation and time-of-flight phase
shifts need to be taken into account. The geometry of the photon
emission is usually expressed as $(\zeta, \Phi)$, where $\zeta$ is
the polar angle from the rotation axis and $\Phi$ is the phase of
rotation of the star. Here we consider the photon emissions from
two poles of pulsar magnetosphere and the photon emission
geometries in the inertial observer's frame (IOF).

The particle motion in the IOF is described by  \citep{TCC07}
\begin{equation} {\bf
n}=[\beta_0\cos\varphi(r)]{\bf b}+[\beta_0{\rm sin}\varphi(r)]{\bf
b_{\bot }}+\beta_{\rm co}{\bf e_\phi},
\end{equation}
where $\varphi(r)$ is the pitch angle. The third term represent
corotation with the star, $\beta_{\rm co}=\rho\Omega/c.$ The
quantity $\beta_0$ at each point is determined by the condition that
${\bf |n|}=1$. The unit vector ${\bf b_\bot}$ perpendicular to the
magnetic field line is
\begin{equation}
{\bf b_\bot}\equiv\pm[({\rm cos}\delta\phi){\bf k}+({\rm
sin}\delta\phi){\bf k\times b}],
\end{equation}
where $\pm$ corresponds to the gyration of the positrons (+) or
electrons (-), $\delta\phi$ refers to the phase of the gyration, and
${\bf k = (b\cdot\nabla)b/|(b\cdot\nabla)b|}$ represents the unit
vector of the curvature of the magnetic field line. Therefore, the
emission direction $(\zeta, \Phi)$ in the IOF is
\begin{equation}
{\rm cos}\zeta = n_z,
\end{equation}
and
\begin{equation}
\Phi = -\Phi_n-{\bf r\cdot n},
\end{equation}
where $-\Phi_n$ is the azimuthal angle of the emission direction,
and {\bf r} is the emitting location in units of the light radius.

We now consider the emission patterns and the examples are shown
in Fig. \ref{Fig5}. In this figure, the emission patterns with
$\epsilon=0.1$, 0, and -0.1 for $\alpha=50^{\circ}$, $60^{\circ}$,
and $70^{\circ}$ are given. For the magnetic field with the
perturbation field, the emission patterns is different from those
for the pure retarded dipole. The emission pattern for
$\epsilon>0$ appears to be more smooth relative to that of the
pure retarded dipole, but the emission pattern for $\epsilon< 0$
is less smooth (see Fig. \ref{Fig5}). Note that the differences
depend on the value of $\epsilon$.

Before calculating the light curves, we consider the change of the
null charge surface with the phase. In the two-pole caustic model
of \citet{DR03}, photons are emitted uniformly in the gap, and the
gap extending from the star surface to high altitudes is confined
to the last open field lines. \citet{FZ10} made a revised version
of the two-pole caustic model, they pointed out that although
acceleration gaps can extend from the star surface to the light
cylinder along near the last open field lines, the extension of
the gaps along the azimuthal direction is limited because of
photon-photon pair production process. In such gaps, high-energy
photons are emitted uniformly and tangentially to the field lines
but cannot be efficiently produced along these field lines where
the distances to the null charge surface are larger than $\sim0.9$
times of the distance of the light cylinder. In Fig. \ref{Fig6},
we show the changes of the null charge surface with the phase for
$\alpha=50^{\circ}$ but different values of $\epsilon$. It can be
seen from this figure that there are substantially difference
between the null charge surfaces of the pure retarded dipole with
and without the perturbation field, and the difference becomes
larger when $\epsilon$ increases.

\begin{figure}
\epsscale{0.7} \plotone{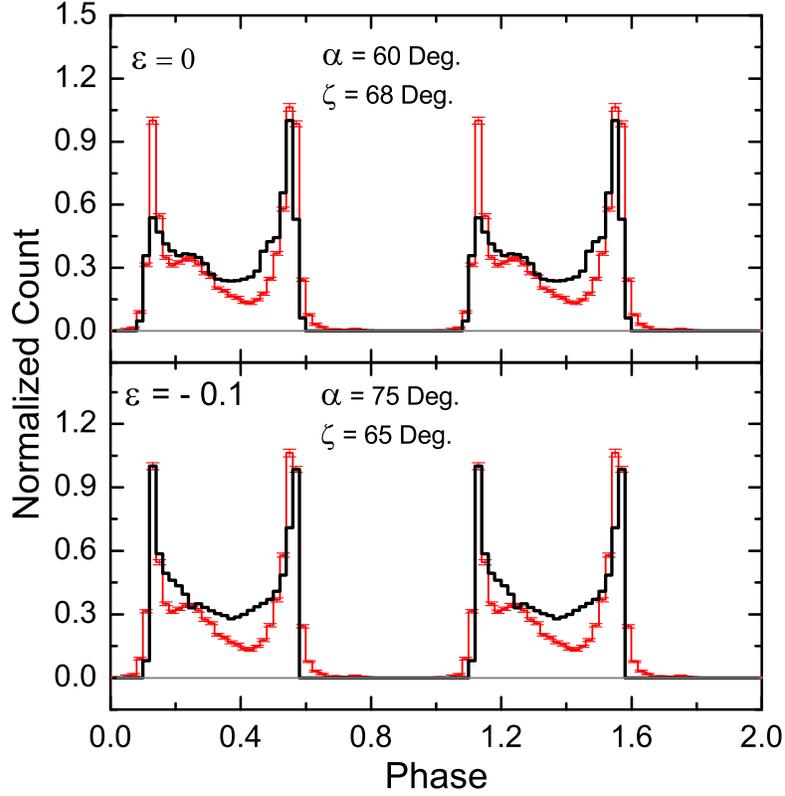} \caption{\label{Fig8} Comparison
of light curves (black solid lines) predicted in two pole outer
gap model with observed light curve (red lines with error bars) in
energy region from 0.1 GeV to 10 GeV of Vela pulsar
\citep{Abdo09}. The predicted light curves in upper and bottom
panels are calculated in the retarded field without any
perturbation and with a perturbation of $\epsilon=-0.1$,
respectively. Note that the predicted light curves come mainly
from the contribution of one pole since the contribution (grey
lines) of another pole can be neglected when the values of
$\alpha$ and $\zeta$ listed in the figure are used.}
%\end{center}
\end{figure}

In calculating the light curves, we use the method given by
\citet{FZ10}, i.e. the emission region is limited in the range of
$r_{\rm null}/R_{\rm L}\le 0.9$ and $\rho_{\rm max}/R_{\rm
L}=0.95$, where $\rho_{\rm max}$ is the distance from the rotation
axis. On the other hand, we use the method given by \citet{DR03}
to smooth the light curves, i.e. the model light curve was
calculated for electrons distributed evenly along the polar cap
rim; the density profile across the rim was assumed to be the
Gaussian function $f (\theta_{\rm m})$ symmetrical about
$\theta_{\rm m} =\theta_{\rm pc}$, with $\sigma=0.025\theta_{\rm
pc}$, where $\theta_{\rm m}$ is the magnetic colatitude of
magnetic field lines' footprints at the star surface, and
$\theta_{\rm pc}\approx (r_{\rm ns}/r_{\rm lc})^{1/2}$ is the
magnetic colatitude of the rim. We show the light curves in
different magnetic inclinations and view angles for
$\epsilon=-0.1$, 0.0, and 0.1 in Fig. \ref{Fig7}, where we have
assumed that the inner boundary of the outer gap is located on the
stellar surface. It can be seen that the change of the light
curves with $\epsilon\neq0$ is remarkable relative to those with
$\epsilon=0$, particularly for larger value of $\epsilon$.
Compared to the light curves with $\epsilon=0$, the light curves
with $\epsilon\neq0$ have following features: (1) peak shapes is
changed although the peak separations are roughly the same, and
(2) the light curves with $\epsilon>0$ are more smoothing but
those with $\epsilon<0$ become more complicated. In Fig.
\ref{Fig8}, we show the light curves predicted in the frame of two
pole outer gap model and compare them with observed one of energy
region from 0.1 GeV to 10 GeV of Vela pulsar \citep{Abdo09}. The
predicted light curve in the upper panel is calculated in the
retarded field without any perturbation, while the lower panel is
calculated in the retarded field with a perturbation of $\epsilon
= - 0.1$. In our calculations for Fig.\ref{Fig8}, the inner
boundary of the outer gap is assumed to be the null charge
surface, the predicted light curves are mainly produced by the
emission of one pole although the contributions of two poles are
included when the values of $\alpha$ and $\zeta$ listed in the
figure are used. Note that the light curves predicted in
\citet{RW10} are based on their one pole outer model given by
\citet{R96}, we can reproduce their result with $\epsilon=0$ in
this one pole outer gap model when we used the parameters (i.e.
$\alpha=72^{\circ}$ and $\xi=64^{\circ}$) listed in Fig. 5 of
\citet{RW10}. Comparing our results with their results we reached
the same conclusion, i.e. that the light curve with $\epsilon<0$
is more consistent with the observed one for the Vela pulsar.

\section{Discussion and Conclusions}
\label{sec:discussion}

Since some currents in the open zone will be created by the
radiating charges accelerated in the gaps and the currents will
induce the magnetic field \citep[e.g.,][]{MH09,RW10}, it is
important to study the structure of the magnetic field with the
perturbation field, emission patterns, and light curves in pulsar
magnetosphere. In this paper, we derive the solution of the static
(retarded) magnetic field with the perturbation field in the
Cartesian coordinates, which can be expressed as the sum of pure
static (retarded) dipolar magnetic field and the perturbation
field in in the Cartesian coordinates. We have confirmed the
reliability of our results by using the solution of the static
magnetic field with the perturbation field (see Fig. \ref{Fig1}),
and then we have used the solution of the retarded magnetic field
with the perturbation field to investigate the emission patterns
and light curves in the pulsar magnetosphere (the parameters of
the Vela pulsar are used). Our results show that the photon
emission pattern and light curves for the retarded magnetic field
with the perturbation field are changed with respective to those
for pure retarded dipolar magnetic field (see Figs. \ref{Fig5} and
\ref{Fig7}).

\citet{RW10} have studied the emission patterns and light curves
of the retarded magnetic field with the perturbation field by
using a highly simplified method. However, our method is totally
different from their method and the results obtained for these two
methods are different (see Figs. \ref{Fig2} and \ref{Fig8}). It
should be noted that a more realistic field to describe pulsar
magnetosphere is the force free field \citep{BS10b}, but such a
filled magnetosphere lacks the acceleration fields required to
produce powerful $\gamma$-rays. \citet{RW10} pointed out that the
force free field can be approximated as a vacuum dipole field
possibly with current-induced perturbation field. Therefore
further works for modelling $\gamma$-ray light curves for
individual pulsars observed by {\it Fermi} are needed in the frame
of more realistic models such as slot gap
\citep[e.g.,][]{HSDF08,HM11} and outer gap
\citep[e.g.,][]{R96,ZC98,Zet04,H08} models in which both physical
and geometrical consideration are taken into account.

\acknowledgments{We thank the anonymous referee for his/her very
constructive comments. This work is partially supported by the
National Natural Science Foundation of China (NSFC 10778702), a
973 Program (2009CB824800), and Yunnan Province under a grant 2009
OC. }

\appendix{\hspace{6cm}\bf{Appendix}}

\section{The Static Dipole Field}

Three components of the static dipole field [Eq. (\ref{BSDfield})]
in Cartesian coordinates are
\begin{eqnarray}
B^{\rm (d)}_x&=&\frac{3}{r^5}(x^2m_x+xym_y+zxm_z)-\frac{m_x}{r^3}\;\nonumber\\
B^{\rm (d)}_y&=&\frac{3}{r^5}(yxm_x+y^2m_y+yzm_z)-\frac{m_y}{r^3}\;\\
B^{\rm
(d)}_z&=&\frac{3}{r^5}(zxm_x+zym_y+z^2m_z)-\frac{m_z}{r^3}\nonumber\;.
\end{eqnarray}
From this set of equations, we can derive three components ($m_x$,
$m_y$, $m_z$) of magnetic moment $\bf{m}$ which is given in Eq.
(3). After including the perturbation field $\bf {B^{(1)}}$, the
magnetic moment $\bf{m}$ is changed to $\bf{m'}=\bf{m}+\epsilon
\bf{m}_{\rm p}$ and the form of Eq. (3) is still valid.
\citet{MH09} derived the analytic expressions of Eq.
(\ref{Bfield}) in magnetic polar coordinates for the static
dipole, which are
\begin{eqnarray}
B_{r} &=& {{B_0^{\rm (d)}}\over {\eta ^3}}~\left(  \cos \theta +
\epsilon ~\chi~{s\over {1-s^2}}  \right) \;,\nonumber\\
B_{\theta } &=& {{B_0^{\rm (d)}}\over {\eta ^3}}~\left[  {1\over
2}~\sin \theta + {{\epsilon ~\chi}\over {1-s^2}}~\left( {{\partial
s}\over {\partial \theta }} + {1\over {\sqrt{1-s^2}}}~{1\over {\sin
\theta}}~{{\partial s}\over {\partial \phi }}  \right) \right]\;,\\
B_{\phi }&=& {{B_0^{\rm (d)}}\over {\eta ^3}}~{{\epsilon ~\chi}\over
{1-s^2}}~\left( {1\over {\sin \theta }}~{{\partial s}\over {\partial
\phi }} - {1\over {\sqrt{1-s^2}}}~{{\partial s}\over {\partial
\theta }}  \right)\nonumber\;,
\end{eqnarray}
where $\eta=r/R$, $R$ is NS radius; $\chi=r/R_{\rm L}$ and $R_{\rm
L}$ is the light cylinder radius;
$s=\cos\alpha\cos\theta+\sin\alpha\sin\theta\cos\phi$ and $\alpha$
is the angle between the rotation axis and magnetic axis. Since the
magnetic field in Eq. (A2) are expressed in magnetic polar
coordinates, the transformation between magnetic polar coordinates
and Cartesian coordinates is
\begin{eqnarray}
{\bf e_r}&=&{\rm sin}\theta{\rm cos}\phi\ {\bf \hat x_m}+{\rm
sin}\theta{\rm sin}\phi\ {\bf \hat y_m}+{\rm cos}\theta\ {\bf \hat
z_m}\;,\nonumber\\
{\bf e_\theta}&=&{\rm sin}(\theta+\pi/2){\rm cos}\phi\ {\bf \hat
x_m}+{\rm sin}(\theta+\pi/2){\rm sin}\phi\ {\bf \hat y_m}+{\rm
cos}(\theta+\pi/2)\ {\bf \hat z_m}\;,\\
{\bf e_\phi}&=&{\rm cos}(\phi+\pi/2)\ {\bf \hat x_m}+{\rm
sin}(\phi+\pi/2)\ {\bf \hat y_m}\nonumber\;,
\end{eqnarray}
where ${\bf \hat z_m}$ is parallel to the magnetic axis. Therefore,
the components of the magnetic field in the Cartesian coordinates
are given by
\begin{eqnarray}
  B_{x_m} &=&\sin\theta\cos\phi B_r+ \sin(\theta+\pi/2)\cos\phi B_{\theta}+ \sin(\theta+\pi/2)B_{\phi}\;, \nonumber\\
  B_{y_m} &=&\sin\theta\sin\phi B_r+ \sin(\theta+\pi/2)\sin\phi B_{\theta}+ \sin(\phi+\pi/2)B_{\phi}\;,\\
  B_{z_m} &=& \cos\theta B_r+ \cos(\theta+\pi/2)B_{\theta} \;.\nonumber
\end{eqnarray}
In principle, converting above components of the magnetic field in
the magnetic polar coordinates into those in the Cartesian
coordinates where the spin axis of the dipole is assumed to be along
the z-axis, we can obtain the expressions of the magnetic field (Eq.
(1)) in the Cartesian coordinates. However, since the perturbation
field $\bf {B^{(1)}}$ will make the magnetic axis {\bf to} curve, we use
$(\alpha_0,~\phi_0)$ to describe the deflection of the magnetic axis
in both  magnetic inclination and azimuthal directions. Through
fitting the results of numerical simulation, we find that $\alpha_0$
and $\phi_0$ are approximated as
\begin{eqnarray}
  \alpha_0 &=&\chi_1\frac{\epsilon r_{xy}}{0.01R_{L}},\ \phi_0=\chi_2 \frac{\epsilon r_{xy}}{0.01R_{L}}\;,\\
  \chi_1&=&-0.02391-0.07578\alpha+0.10575\alpha^2-0.03825\alpha^3\;,\nonumber\\
  \chi_2&=& 0.00189-0.08582e^{-2.49738\alpha}\;, \nonumber
\end{eqnarray}
where $r_{xy}$ is the distance to the spin axis. After eliminating
the influence of curved magnetic axis, we can use following
expressions to obtain the expressions of the magnetic field ( Eq.
(1)) in the Cartesian coordinates:
\begin{eqnarray}
  B_{x} &=& \cos(\alpha)B_{x_m}+\sin(\alpha)B_{z_m}\;, \nonumber\\
  B_{y} &=& B_{y_m} \;,\\
  B_{z} &=& \cos(\alpha)B_{z_m}-\sin(\alpha)B_{x_m}\;. \nonumber
\end{eqnarray}
Replacing $B^{(d)}_x$, $B^{(d)}_y$, and $B^{(d)}_z$ into $B_x$,
$B_y$, and $B_z$ in Eq. (3), three components of magnetic moment
$\bf{m'}$ of the magnetic field given by Eq. (1) can be obtained
as follow:
\begin{eqnarray}
m'_x&=&\frac{r}{2}\left[(3x^2-2r^2)B_x+3xyB_y+3xzB_z\right]\;,\nonumber\\
m'_y&=&\frac{r}{2}\left[3yxB_x+(3y^2-2r^2)B_y+yzB_z\right]\;,\\
m'_z&=&\frac{r}{2}\left[3xzB_x+3yzB_y+(3z^2-2r^2)B_z\right]\;,\nonumber
\end{eqnarray}
where the perturbation magnetic moment $\bf{m}_p$ can be expressed
as
\begin{eqnarray}\label{msp}
m_{px}&=&\frac{r}{2}\left[(3x^2-2r^2) B^{(1)}_{x}+3xy B^{(1)}_{y}+3xz B^{(1)}_{z}\right]\;,\nonumber\\
m_{py}&=&\frac{r}{2}\left[3yx B^{(1)}_{x}+(3y^2-2r^2) B^{(1)}_{y}+yz B^{(1)}_{z}\right]\;,\\
m_{pz}&=&\frac{r}{2}\left[3xz B^{(1)}_{x}+3yz B^{(1)}_{y}+(3z^2-2r^2)
B^{(1)}_{z}\right]\;.\nonumber
\end{eqnarray}

\section{The Retarded Dipole Field}

\subsection{Retarded Magnetic Moment}

For an inclined dipole without the perturbation field of constant
magnitude, rotating with angular frequency $\Omega$ about the
$\hat{z}$-axis, the magnetic moment is expressed as
\begin{equation}
{\bf m^r}=m({\rm sin}\alpha{\rm cos}(\Omega t)\ {\bf \hat x}+ {\rm
sin}\alpha{\rm sin}(\Omega t)\ {\bf \hat y}+{\rm cos}\alpha\ {\bf
\hat z})\;,
\end{equation}
i.e. the magnetic moment for the static dipole rotates along the
rotational direction by $\Omega t=(r_{0}-r)/R_L$. For the same
reason, three components for rotating perturbation magnetic moment
$\bf m^r_p$ are given by
\begin{eqnarray}\label{mpr}
m^r_{px}&=& m_{px}\cos(\Omega t)+m_{py}\sin(\Omega t)\;,\nonumber\\
m^r_{py}&=& m_{px}\sin(\Omega t)+m_{py}\cos(\Omega t)\;,\\
m^r_{pz}&=& m_{pz}\;.
\end{eqnarray}
Therefore, the magnetic moment which includes the perturbation
field is given by
\begin{equation}
\bf{m^{'r}}=\bf{m^r} + \epsilon \bf{m}^r_p\;.
\end{equation}

\subsection{Magnetic field with the Perturbation Field }

For the pure retarded dipole, the magnetic field is expressed as
\citep{CRZ00}
\begin{equation}
{\bf B}^{r}=\hat {\bf r} \left[ \hat{\bf r} \cdot \left(
\frac{3{\bf
m^{r}}}{r^3}+\frac{3{\bf\dot{m}^{r}}}{cr^2}+\frac{{\bf
\ddot{m}^{r} }}{c^2r} \right) \right] -\left( \frac{{\bf
m}^{r}}{r^3}+\frac{{\bf\dot{m}^{r}}}{cr^2}+\frac{{\bf
\ddot{m}^{r} }}{c^2r} \right)\;, \label{Bfield1}
\end{equation}
where $r$ is radial distance, $\hat{\bf r}$ is the radial unit
vector, and $c$ is the light speed. Replacing $\bf{m}^{r}$ into
$\bf{m}^{'r}=\bf{m}^{r} +\epsilon\bf{m}^{r}_{p}$ in above
equation, the magnetic field with the perturbation field is given
by
\begin{equation}
{\bf B^{'r}}={\bf B^{r}}+\epsilon {\bf B^{r}_{p}}\;,
\end{equation}
where the rotating perturbation field ${\bf B^{r}_{p}}$ is given
by
\begin{equation}
{\bf B^r_p}=\hat {\bf r} \left[ \hat{\bf r} \cdot \left(
\frac{3{\bf m^r_p}}{r^3}+\frac{3{\bf\dot{m^r_p}}}{cr^2}+\frac{{\bf
\ddot{m^r_p} }}{c^2r} \right) \right] -\left( \frac{{\bf
m^r_p}}{r^3}+\frac{{\bf\dot{m^r_p}}}{cr^2}+\frac{{\bf \ddot{m^r_p}
}}{c^2r} \right)\;.
\end{equation}
The three components in the Cartesian coordinates can expressed as
\begin{eqnarray}
B^r_{px}&=& {\bf \hat x}\cdot{\bf B^r_p}=\frac{1}{r^5}\left[{\rm
C_x}{\rm cos}(\frac{R-r}{R_L})+{\rm
D_x}{\rm sin}(\frac{R-r}{R_L})+3xzm_{pz}\right]\;,\nonumber\\
B^r_{py}&=& {\bf \hat y}\cdot{\bf B^r_p}=\frac{1}{r^5}\left[{\rm
C_y}{\rm cos}(\frac{R-r}{R_L})+{\rm D_y}{\rm
sin}(\frac{R-r}{R_L})+3yzm_{pz}\right]\;,\nonumber\\
B^r_{pz}&=& {\bf \hat z}\cdot{\bf B^r_p}=\frac{1}{r^5}\left[{\rm
C_z}{\rm cos}(\frac{R-r}{R_L})+{\rm D_z}{\rm
sin}(\frac{R-r}{R_L})+(r^2-3z^2)m_{pz}\right]\;,
\end{eqnarray}
where
\begin{eqnarray}
{\rm
C_x}=(r^2m_{px}-xym_{py}-x^2m_{px})\frac{r^2}{R_L^2}-(r^2m_{py}-
3xym_{px}-3x^2m_{py})\frac{r}{R_L} -\nonumber\\
(r^2m_{px}-3xym_{py}-3x^2m_{px}),\nonumber\\
{\rm
D_x}=(r^2m_{py}-xym_{px}-x^2m_{py})\frac{r^2}{R_L^2}+(r^2m_{px}-
3xym_{py}-3x^2m_{px})\frac{r}{R_L}
-\nonumber\\(r^2m_{py}-3xym_{px}-3x^2m_{py}),\nonumber\\
{\rm
C_y}=(r^2m_{py}-xym_{px}-y^2m_{py})\frac{r^2}{R_L^2}-(r^2m_{px}-3xym_{py}-
3y^2m_{px})\frac{r}{R_L}-\nonumber\\(r^2m_{py}-3xym_{px}-3y^2m_{py}),\nonumber\\
{\rm
D_y}=(r^2m_{px}-xym_{py}-y^2m_{px})\frac{r^2}{R_L^2}+(r^2m_{py}-3xym_{px}-
3y^2m_{py})\frac{r}{R_L}-\nonumber\\(r^2m_{px}-3xym_{py}-3y^2m_{px}),\nonumber\\
{\rm C_z}=(-xzm_{px}-yzm_{py})\frac{r^2}{R_L^2}+(3xzm_{py}+
3yzm_{px})\frac{r}{R_L}+(3xzm_{px}+3yzm_{py}),\nonumber\\
{\rm D_z}=(-xzm_{py}-yzm_{px})\frac{r^2}{R_L^2}-(3xzm_{px}+
3yzm_{py})\frac{r}{R_L}+(3xzm_{py}+3yzm_{px})\nonumber.
\end{eqnarray}


\begin{thebibliography}{99}

\bibitem[Abdo et al. (2009)]{Abdo09}
Abdo, A. A. et al.  2009, ApJ, 696, 1084

\bibitem[Abdo et al.(2010)]{Abdo10}
Abdo, A. A. et al.2010, ApJS, 187, 460

\bibitem[Bai \& Spitkovsky(2010a)]{BS10a}
Bai, X. -N. \& Spitkovsky, A. 2010a, ApJ, 715, 1282

\bibitem[Bai \& Spitkovsky(2010b)]{BS10b}
Bai, X. -N. \& Spitkovsky, A. 2010b, ApJ, 715, 1270

\bibitem[Cheng et al.(2000)]{CRZ00}
Cheng, K. S., Ruderman, M. A., \& Zhang, L. 2000 ApJ, 537, 964

\bibitem[Deutsch(1955)]{Deutsch55}
Deutsch, Arnim J.1955, AnAp, 18, 1D

\bibitem[Dyks \& Rudak(2003)]{DR03}
Dyks, J., \& Rudak, B. 2003, \apj, 598, 1201
\bibitem[Dyks et al. (2004)]{DHR04}
Dyks, J., Harding, A. K., Rudak, B. 2004, \apj, 606, 1124

\bibitem[Fang \& Zhang(2010)]{FZ10}
Fang, J., \& Zhang, L. 2010, \apj, 709, 605

\bibitem[Harding et al.(2008)]{HSDF08}
Harding, A. K., Stern, J. V., Dyks, J., Frackowiak, M. 2008, ApJ,
680, 1378

\bibitem[Harding \& Muslimov (2011)]{HM11}
Harding, A. K., Muslimov, A. G. 2011, ApJ, 726, L10

\bibitem[Hirotani (2008)]{H08}
Hirotani, K. 2008, \apj, 688, L25

\bibitem[Li \& Zhang(2010)]{LZ10}
Li, X. \& Zhang, L. 2010, ApJ, 725, 2225

\bibitem[Muslimov \& Harding(2009)]{MH09}
Muslimov, A. G. \& Harding, A. K. 2009, ApJ, 692, 140

\bibitem[Romani \& Yadigaroglu, I.-A.(1995)]{RY95}
Romani, R. W. \& Yadigaroglu, I.-A., 1995, \apj, 438, 314

\bibitem[Romani (1996)]{R96}
Romani, R. W. 1996, \apj, 470, 469

\bibitem[Romani \& Watters(2010)]{RW10}
Romani, R. W., \& Watters K. P. 2010, \apj, 714, 810

\bibitem[Takata et al.(2007)]{TCC07}
Takata, J., Chang, H.-K., \& Cheng, K. S. 2007, ApJ, 656, 1044

\bibitem[Wang et al.(2011)]{WTC11}
Wang, Y., Takata, J., Chang, H.-K., \& Cheng, K. S. 2011, MNRAS,
tmp 575, arXiv:1102.4474

\bibitem[Tang et al. (2008)]{Tet08}
Tang, A. P. S., Takata, T., Jia, J. J., \& Cheng, K. S. 2008, ApJ,
676, 562

\bibitem [Watters et al. (2009)]{W09}
Watters K. P., Romani, R. W. Weltevrede, P. \& Johnston, S. 2009
ApJ, 695, 1289

\bibitem[Zhang \& Cheng(1998)]{ZC98}
Zhang, L., \& Cheng, K. S. 1998, MNRAS, 294, 177

\bibitem[Zhang \& Cheng(2000)]{ZC00}
Zhang, L., \& Cheng, K. S. 2000, A\&A, 363, 575

\bibitem[Zhang \& Cheng(2001)]{ZC01}
Zhang, L., \& Cheng, K. S. 2001, MNRAS, 320, 477

\bibitem[Zhang \& Cheng (2002)]{ZC02}
Zhang, L. \& Cheng, K. S. 2002, ApJ, 569, 872

\bibitem[Zhang et al. (2004)]{Zet04}
Zhang, L., Cheng, K. S., Jiang, Z. J., \& Leung, P. 2004, ApJ,
604, 317

\bibitem[Zhang \& Li(2009)]{ZL09}
Zhang, L.,  \& Li, X. 2009, ApJ, 707, L169


\end{thebibliography}
\end{document}